\documentclass[a4paper,12pt,texlive=2016]{article}

\RequirePackage{graphicx}
\RequirePackage{subfig}
\RequirePackage{mhchem}
\RequirePackage{amssymb}

\newcommand{\TeV}{\ensuremath{\text{Te\kern -0.1em V}}}
\newcommand{\GeV}{\ensuremath{\text{Ge\kern -0.1em V}}}
\newcommand{\pt}{\ensuremath{p_{\text{T}}}}
\newcommand{\met}{\ensuremath{E_{\text{T}}^{\text{miss}}}}

\title{Study of interference effects in the search for flavour-changing neutral current interactions involving the top quark and a photon or a $Z$~boson at the LHC}

\author{Maura~Barros$^1$, Nuno~Filipe~Castro$^{1,2}$, Johannes~Erdmann$^3$,\\Gregor~Ge{\ss}ner$^3$, Kevin~Kr\"oninger$^3$, Salvatore~La~Cagnina$^3$,\\Ana~Peixoto$^1$
}

\date{\small
  $^1$ Laborat\'orio de Instrumenta\c{c}\~{a}o e F\'isica Experimental de Part\'iculas (LIP), Universidade do Minho, Braga, Portugal\\
  $^2$ Departamento de F\'isica, Escola de Ci\^encias, Universidade do Minho, Braga, Portugal\\
  $^3$ Technische Universit\"at Dortmund, Dortmund, Germany
}

\begin{document}

\maketitle

\begin{abstract}
Flavour-changing neutral-current interactions of the top quark can be searched for in top-quark pair production with one top quark decaying to an up-type quark and a neutral boson, and they can be searched for in the single production of a top quark in association with such a boson. Both processes interfere if an additional up-type quark is produced in the case of single production. The impact of these interference effects on searches for flavour-changing neutral currents at the LHC is studied for the case where the neutral boson is a photon or a $Z$ boson. Interference effects are found to be smaller than variations of the renormalisation and factorisation scales.

\end{abstract}

\section{Introduction}
\label{sec:introduction}
In the Standard Model~(SM), flavour-changing neutral currents~(FCNCs) are forbidden at tree level and highly suppressed at loop level. For FCNCs that involve the top quark, the SM predictions for top-quark branching ratios range from $2 \times 10^{-17}$ for the decay $t \rightarrow uH$ to $4.6 \times 10^{-12}$ for the decay $t \rightarrow cg$~\cite{AguilarSaavedra:2004wm}. Several extensions of the SM predict branching ratios that are much larger and may be accessible at the Large Hadron Collider~(LHC)~\cite{AguilarSaavedra:2004wm}. An observation of FCNCs in the top-quark sector would hence be a clear sign of physics beyond the SM. Deviations from the SM predictions in these searches can be described in a model-independent way using an effective-field theory~(EFT) approach~\cite{AguilarSaavedra:2008zc,Zhang:2010dr}. In lowest order, these deviations are described by dimension-six operators, suppressed by the square of the new-physics scale.

\begin{figure}
  \centering
  \subfloat[]{\includegraphics[scale=0.5]{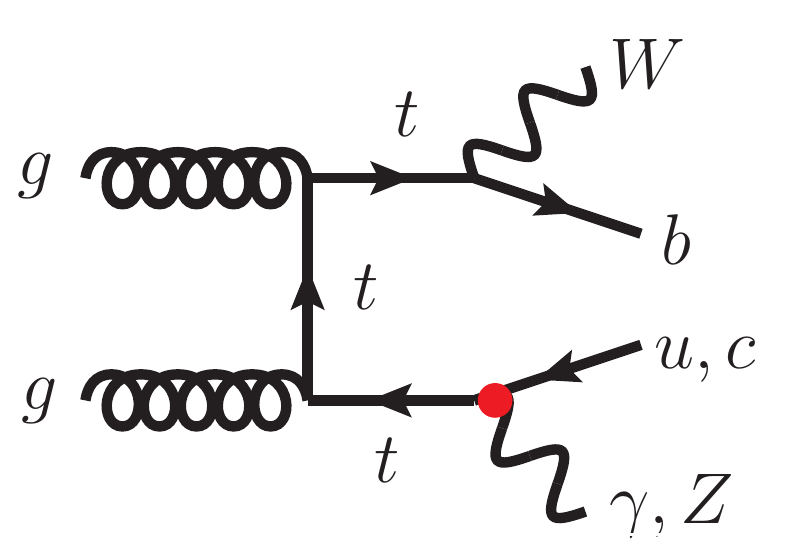}}
  \subfloat[]{\raisebox{5mm}{{\includegraphics[scale=0.5]{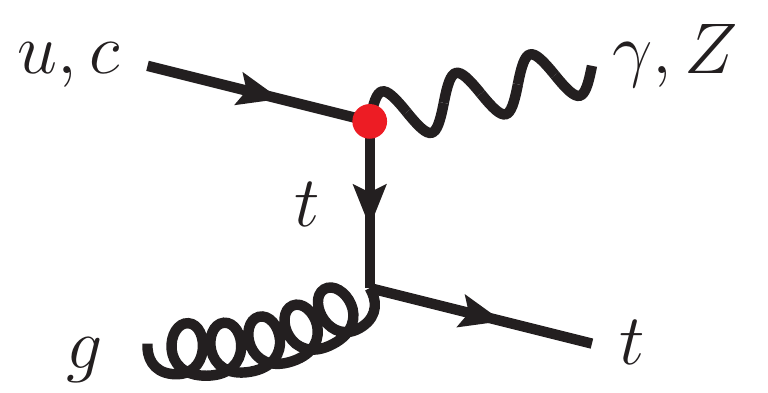}}}}
  \caption{Example of Feynman diagrams (a) for top-quark pair production with an FCNC top-quark decay to a photon or a $Z$ boson and (b) for the associated production of a single top quark together with a photon or a $Z$ boson via an FCNC interaction. The FCNC vertex is marked with a red dot.}
\label{fig:feynman}
\end{figure}

Top-quark FCNCs can be searched for in top-quark pairs where one top quark decays into a $W$ boson and a $b$-quark and the other top quark decays via an FCNC, as shown in Figure~\ref{fig:feynman}~(a). They can also be searched for in the production via an FCNC, as presented in Figure~\ref{fig:feynman}~(b). The former process is called ``the decay process'' in the following, and the latter process is called ``the production process''.

Interference effects have shown to play an important role in several studies of the top-quark sector. The interference of two SM processes, top-quark pair production and the production of a single top quark in association with a $W$ boson, was studied in Ref.~\cite{Aaboud:2018bir}, constraining interference models for these processes. In searches for new particles, considering interference effects has also proven to be important. For example, in the search for new scalars and pseudoscalars that decay to a top-quark pair, interference effects significantly alter the shape of the narrow signal resonance on the falling background to a peak--dip structure~\cite{Aaboud:2017hnm}. A similar effect was observed in the search for fermionic top-quark partners that decay to a $W$ boson and a $b$-quark~\cite{Aaboud:2018ifs}.

Searches for top-quark FCNCs have focused either on the decay process, for example in Refs.~\cite{Aaboud:2018nyl,CMS-PAS-TOP-17-017}, or on the production process, for example in Refs.~\cite{Aad:2019pxo,Sirunyan:2017kkr}. However, both processes interfere if at least one additional up-type quark is produced in the production process. These interference effects may have an impact on the interpretation of a potential observation of an FCNC signal and on the exclusion limits that are set by current and future searches.

The goal of this study is to quantify whether these interference effects should be considered in the experimental searches at the LHC or, on the contrary, they can be safely neglected, as it has been done to date. We have limited ourselves to the study of interference effects in top-quark FCNC processes with either a photon or a $Z$ boson. Processes with a Higgs boson should be studied separately. Processes with a gluon are experimentally probed via the process $gq \rightarrow t$~\cite{Aad:2015gea,Khachatryan:2016sib}, where $q$ is either an up or a charm quark, i.e. not via top-quark decays.

\section{Monte Carlo samples}

Monte Carlo (MC) samples were generated for proton--proton collisions at $\sqrt{s} = 13~\TeV$ with \textsc{MadGraph5\_} \textsc{aMC@NLO}~\cite{Alwall:2014hca} at leading order using the TopFCNC \cite{Degrande:2014tta,Durieux:2014xla} UFO~\cite{Degrande:2011ua} model for the production and decay processes. In addition, a sample that includes both processes and their interference, called ``the total process'', was generated. For all samples, dynamic factorisation and renormalisation scales were used as well as the NNPDF2.3LO PDF set~\cite{Ball:2012cx}. In the following, the generation of the processes with a photon is discussed. The samples for the processes with a $Z$ boson were generated analogously. The quark $q$ can be either an up quark or a charm quark.

The decay process was generated by $pp \rightarrow t\bar{t}$ with $t\bar{t} \rightarrow W^+b\gamma\bar{q}$ or $t\bar{t} \rightarrow W^-\bar{b}\gamma q$.

The production process was generated by $pp \rightarrow t\gamma$ (or $\bar{t}\gamma$) and adding the processes with an extra quark, anti-quark or gluon, $j$, i.e. $pp \rightarrow t\gamma j$ (or $pp \rightarrow \bar{t}\gamma j$). Diagrams with an intermediate $\bar{t}$ ($t$) were excluded in the case of $t\gamma(j)$ ($\bar{t}\gamma(j)$) production, because these diagrams are considered as part of the decay process. In all cases, the (anti-)top quarks were decayed to $W^+b$ ($W^-\bar{b}$) using \textsc{MadSpin}~\cite{Artoisenet:2012st}. If $j = q$, i.e. the extra quark corresponds to the up-type quark that couples via an FCNC to the top quark, interference of the decay and production processes occurs.

The total process was generated by $pp \rightarrow t\gamma$ (or $\bar{t}\gamma$) and by adding the processes $pp \rightarrow t\gamma j$ (or $\bar{t}\gamma j$). Again, the (anti-)top quarks were decayed to $W^+b$ ($W^-\bar{b}$) using \textsc{MadSpin}.

All $W$ bosons were decayed to an electron or muon and the corresponding neutrino, and all $Z$ bosons were decayed to either two electrons or two muons using \textsc{MadSpin}. A minimum $\pt$ of 20~\GeV\ was required for every final-state parton in the generation with \textsc{MadGraph5\_aMC@NLO}.

\textsc{Pythia}~8.2~\cite{Sjostrand:2014zea} was used for parton showering and hadronisation. For all processes, events were matched using the MLM procedure~\cite{Mangano:2006rw} using a $k_t$ value between partons of 30~\GeV. Even though not necessary for the decay process, we also apply MLM matching to this process for a consistent treatment of all processes. Detector effects were simulated with \textsc{Delphes}~3 \cite{deFavereau:2013fsa} using the default detector card, tuned to match the CMS detector parameters. An overview of the number of events generated for each sample and the number of contributing Feynman diagrams is shown in Table~\ref{tab:samples}.

The FCNC operators may be left- or right-handed and they may couple the top quark to the up quark or to the charm quark~\cite{AguilarSaavedra:2008zc}. For each of these four possibilities, samples were produced with the strength of one operator fixed to a benchmark value and that of the other operators set to zero. It was verified that the choice of the benchmark value used does not alter the kinematics of the processes unless the value would be chosen so large that it modifies the intrinsic width of the top quark significantly\footnote{Such large values of the strengths of the operators are, however, excluded by searches for FCNC processes at the LHC~\cite{Aaboud:2018nyl,CMS-PAS-TOP-17-017,Aad:2019pxo,Sirunyan:2017kkr}.}. Uncertainties were evaluated by generating additional samples with fixed renormalisation and factorisation scales set to the top-quark mass and by varying the scales by factors of two. The variations in the fixed-scale samples were used as relative uncertainties for the nominal samples, which were generated with a dynamic scale defined as the transverse mass of the system after $k_t$ clustering of the final-state particles.

\begin{table}
  \centering
  \begin{tabular}{lll}
    \hline\noalign{\smallskip}
    Process & No. of events & No. of Feynman diagrams \\
    \noalign{\smallskip}\hline\noalign{\smallskip}
    Production & 500,000 & 84 \\
    Decay & 500,000 & 16 \\
    Total & 500,000 & 100 \\
    \noalign{\smallskip}\hline\noalign{\smallskip}
  \end{tabular}
  \caption{Number of events generated for each process and the number of Feynman diagrams included in the generation.}
  \label{tab:samples}
\end{table}

\section{Results}

The cross sections for the different processes are presented in Table~\ref{tab:xsecs}, showing that the cross section for the sum of the decay and production processes is very close to the cross section of the total process. In order to quantify the impact of considering or neglecting interference effects, the sample generated for the total process was compared to the cross-section weighted sum of the samples for the production and decay processes. Only results for the left-handed coupling of the top quark to the up quark are shown as an example. The conclusions for the right-handed coupling and for the left- and right-handed couplings to the charm quark are the same as for the example shown. Results at parton and at detector level are discussed.

\begin{table}
  \centering
  \begin{tabular}{llll}
    \hline\noalign{\smallskip}
    Boson & Decay   & Production & Total \\
          & process & process    & process \\
    \noalign{\smallskip}\hline\noalign{\smallskip}
    $\gamma$ & $6\pm 3$ fb & $13\pm 5$ fb & $20\pm 8$ fb\\
    $Z$      & $24\pm 11$ fb & $85\pm 30$ fb & $110\pm 40$ fb\\
    \noalign{\smallskip}\hline\noalign{\smallskip}
  \end{tabular}
  \caption{Cross sections for the different processes including the uncertainty from scale variations.}
  \label{tab:xsecs}
\end{table}

\subsection{Parton level}

\begin{figure}
  \centering
  \subfloat[]{\includegraphics[width=0.4\textwidth]{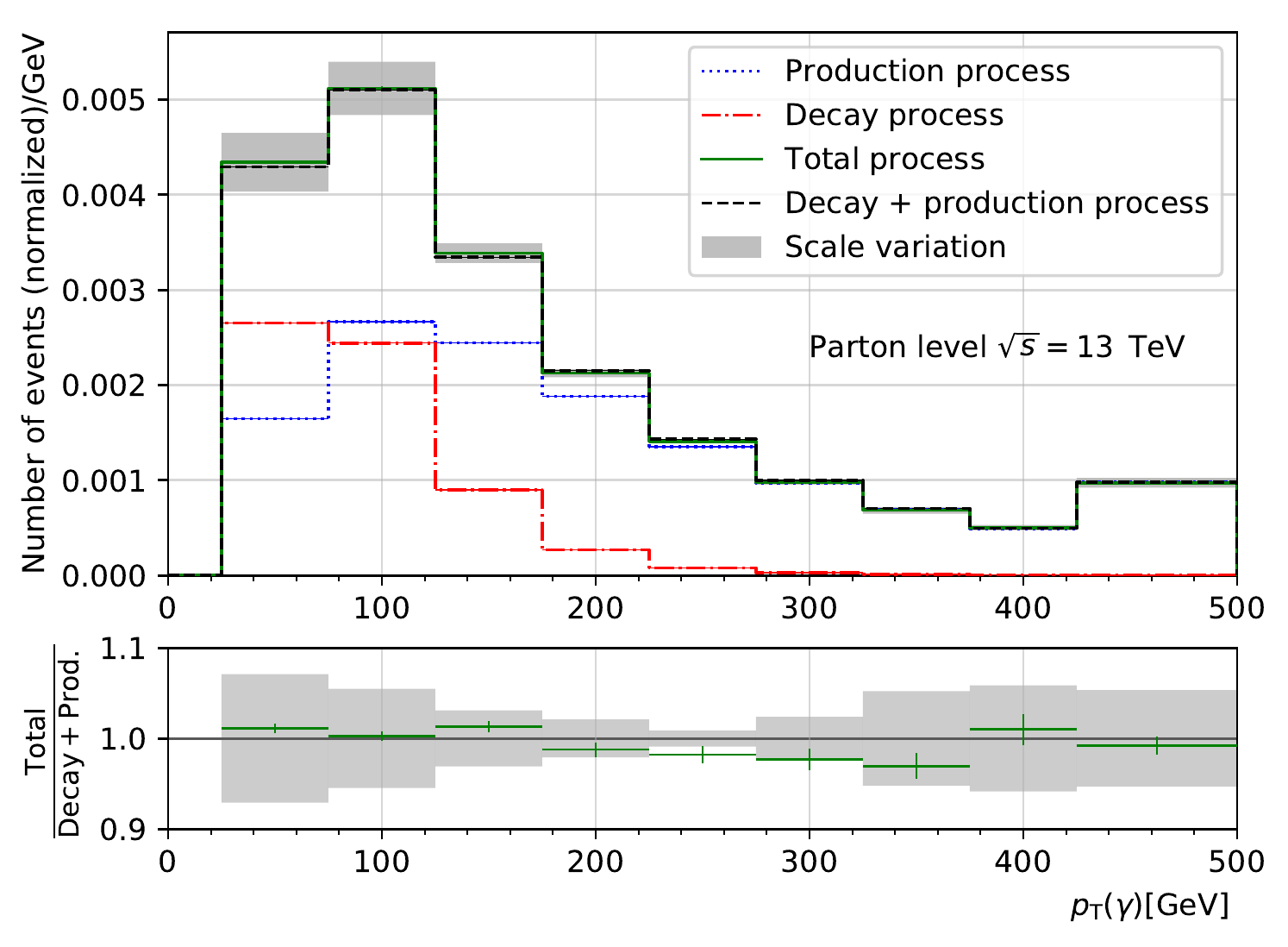}}%journal: \\
  \quad %journal: remove
  \subfloat[]{\includegraphics[width=0.4\textwidth]{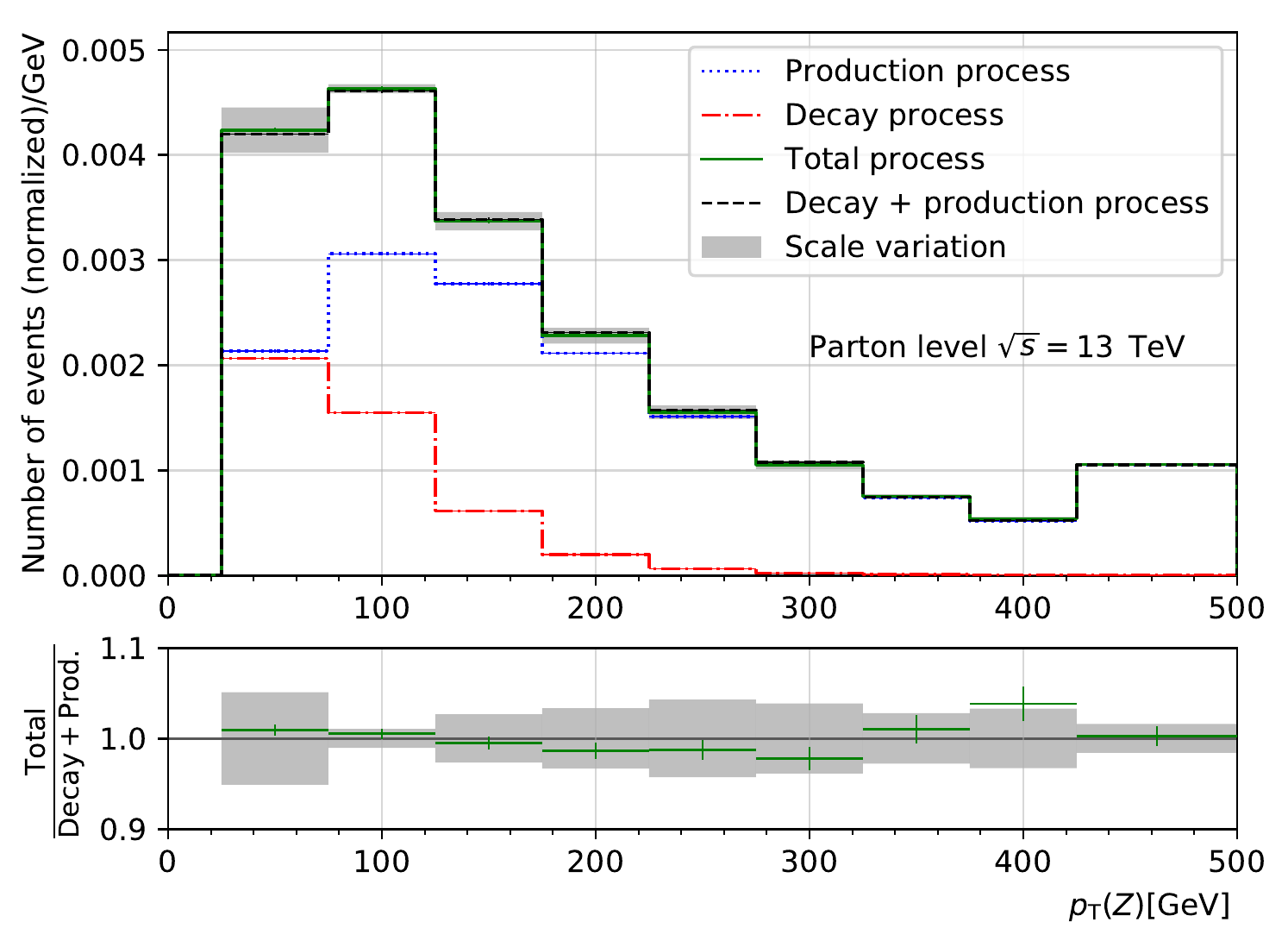}}\\
  \caption{The (a) photon and (b) $Z$ boson transverse momentum, respectively, for the total process compared to the sum of the production and decay processes at parton level. The distributions for the total process and for the sum of the production and the decay process are each normalised to unity. The ratio of the total process and the sum of the processes is also shown. The uncertainty in the total process due to variations of the renormalisation and factorisation scales is shown as a band. The statistical uncertainty due to the limited size of the samples is shown as a green error bar.}
  \label{fig:parton}
\end{figure}

Kinematic distributions of the different final-state parti-cles---i.e. the photon~/ $Z$ boson\footnote{The kinematics of the $Z$ boson were considered before final-state radiation, i.e. directly at the flavour-changing vertex.}, the $b$ quark, the top quark, the $W$ boson, and the highest-\pt\ up-type quark---were studied at parton level and no large effects were seen. As examples, the normalised distributions of the transverse momentum (\pt) of the photon / $Z$ boson from the total process and from the sum of the production and decay processes are shown in Figure~\ref{fig:parton}.

In the phase space where both the production and the decay process contribute and interference effects could hence appear, i.e. \pt\ smaller than approximately 300~\GeV, the difference between the total process and the sum of the two individual processes is small and covered by the scale uncertainties\footnote{For larger \pt, the distributions of the total process and the production process are consistent within statistical uncertainties.}. This implies that effects due to interference are small compared to the systematic uncertainties from scale variations.

\subsection{Detector level}

The photon and $Z$-boson \pt\ were also studied at detector level. A simple event selection was used in order to mimic the selection of a search: In each event at least one photon or $Z$ boson was required, respectively. The photon was required to fulfil $\pt > 25~\GeV$ and $|\eta|<2.5$, where $\eta$ is the pseudorapidity. $Z$ bosons were reconstructed from two electrons or two muons with an invariant mass within a $10~\GeV$ window around the $Z$-boson mass. If several $Z$-boson candidates were found, the one with the mass closest to the $Z$-boson mass was selected. The leptons were required to have opposite electric charge, $\pt > 25~\GeV$ and $|\eta| < 2.5$. Moreover, at least one (additional) electron or muon with the same \pt\ and $\eta$ criteria was required, as well as $\met > 20~\GeV$, at least one $b$-tagged jet and at least one non-$b$-tagged jet, reconstructed with the anti-$k_t$ algorithm~\cite{Cacciari:2008gp} with a radius parameter of $R=0.5$, with $\pt > 25~\GeV$ and $|\eta| < 2.5$.

\begin{figure}
  \centering
  \subfloat[]{\includegraphics[width=0.4\textwidth]{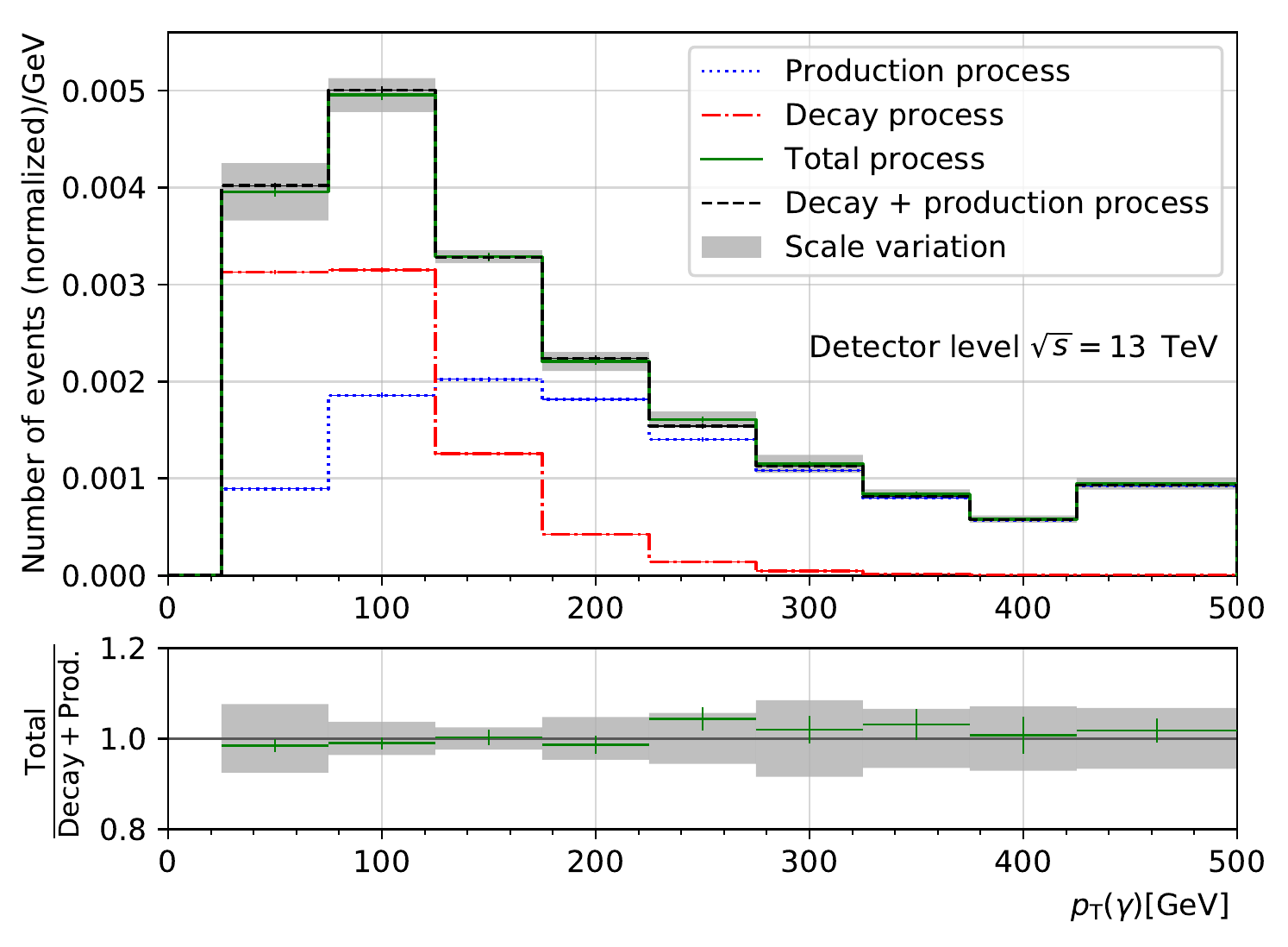}}%journal: \\
  \quad %journal: remove
  \subfloat[]{\includegraphics[width=0.4\textwidth]{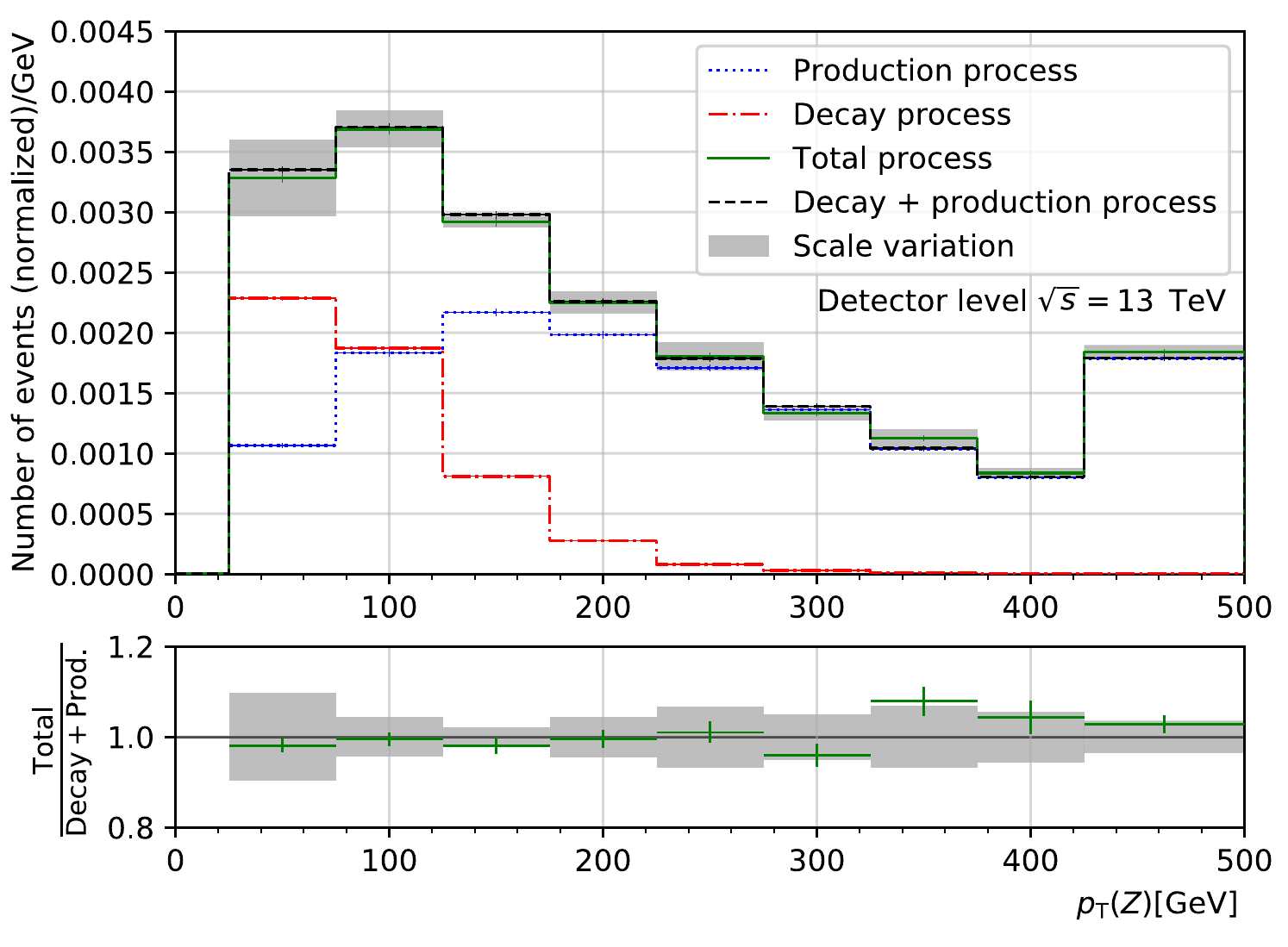}}\\
  \caption{The (a) photon and (b) $Z$ boson candidate transverse momentum, respectively, for the total process compared to the sum of the production and decay processes at detector level. The distributions for the total process and for the sum of the production and the decay process are each normalised to unity. The ratio of the total process and the sum of the processes is also shown. The uncertainty in the total process due to variations of the renormalisation and factorisation scales is shown as a band. The statistical uncertainty due to the limited size of the samples is shown as a green error bar.}
  \label{fig:detector}
\end{figure}

The normalised distributions from the total process and from the sum of the production and decay processes are shown in Figure~\ref{fig:detector}. The same observations and conclusions hold as for parton level: No large interference effects are seen where they could be expected ($\pt \lessapprox 300~\GeV$).

\section{Conclusions}

Interference effects in top-quark processes with flavour-changing neutral currents were studied for proton--proton collisions at 13~\TeV, focusing on interactions that involve a photon or a $Z$ boson. Interference effects were found to be much smaller than changes from variations of the renormalisation and factorisation scales in the leading-order (multileg) samples used for this study. These results indicate that the current practise of neglecting interference effects in searches for top-quark flavour-changing neutral current interactions at the LHC is a viable (and practical) strategy also for the future. However, if flavour-changing neutral currents were observed in such searches, the impact of interference effects on their interpretations should be quantified.

\section*{Acknowledgements}
Open Access funding provided by Projekt DEAL. This work was supported by OE/FCT, Lisboa2020, Compete2020, Portugal 2020 and FEDER through project POCI/01-0145-FEDER-029147, PTDC/FIS-PAR/29147/2017; by\\project CERN/FIS-PAR/0008/2017 (OE/FCT); by grant SFRH/BD/129321/\\2017 (OE/FCT); by the DFG through project KR 4060/7-1; and by the BMBF via FSP-103 through projects 05H15PECAA and 05H19PECA1.

\bibliographystyle{phaip}
\bibliography{FCNCinterference}

\end{document}